\documentclass[11pt,a4paper]{article}
\usepackage{jheppub2}
\usepackage{amssymb}
\usepackage{dsfont}
\usepackage{color}
\usepackage{slashed}

\usepackage{dsfont}

\newcommand{\df}{\Delta_\phi}
\newcommand{\ds}{\Delta_\sigma}
\newcommand{\nmax}{$n_{\mbox{\tiny max}}$}
\newcommand{\mmax}{$m_{\mbox{\tiny max}}$}

\newcommand{\jb}{{\tt JuliBootS}}

\newcommand{\reef}[1]{(\ref{#1})}
\newcommand{\be}{\begin{equation}}
\newcommand{\ee}{\end{equation}}
\newcommand{\bea}{\begin{eqnarray}}
\newcommand{\eea}{\end{eqnarray}}

\newcommand{\mbf}{\mathbf}

\newenvironment{code}
{\begin{flushleft}\begin{quote}\small\ttfamily}
{\end{quote}\end{flushleft}}

\newcommand{\bec}{\begin{code}}
\newcommand{\eec}{\end{code}}

\title{{\tt JuliBootS}: a hands-on guide to the conformal bootstrap.}

\author[a,b]{Miguel F. Paulos}

\affiliation[a]{Department of Physics,
Brown University,
Box 1843,
Providence, RI 02912-1843,
USA}
\affiliation[b]{Theory Division, CERN, Geneva, Switzerland}

\abstract{We introduce {\tt JuliBootS}, a package for numerical conformal bootstrap computations coded in {\tt Julia}. The centre-piece of {\tt JuliBootS} is an implementation of Dantzig's simplex method capable of handling arbitrary precision linear programming problems with continuous search spaces. Current supported features include conformal dimension bounds, OPE bounds, and bootstrap with or without global symmetries. The code is trivially parallelizable on one or multiple machines. We exemplify usage extensively with several real-world applications. In passing we give a pedagogical introduction to the numerical bootstrap methods.
}

\begin{document}
\maketitle

\section{Introduction}

The conformal bootstrap program is the hope that conformal field theories might be so constrained as to be exactly solvable given some minimal set of assumptions. This hope has been largely realized for rational conformal field theories in two dimensions, where the power of complex analysis comes to the rescue under the form of an infinite dimensional extension of the conformal group \cite{Belavin:1984vu}. In higher dimensions, progress was slow in the last 40 years since the original proposal of the program by Polyakov \cite{Polyakov:1974gs} building on the work of Ferrara, Gatto, Grillo and Parisi \cite{Ferrara:1974ny,Ferrara:1974pt,Ferrara:1973yt,Ferrara:1971vh,Ferrara:1974nf,Ferrara:1973vz} . However more recently, thanks to computational advances together with systematic studies of conformal blocks \cite{Dolan2011,DO2,DO1}, the bootstrap program has been revived with great success by Rattazzi, Rychkov, Tonni and Vichi \cite{Rattazzi:2008pe} (a selection of further developments is \cite{El-Showk2014,ElShowk:2012hu,El-Showk2014a,ElShowk:2012ht,Poland:2010wg,Rattazzi:2010gj,Rattazzi:2010yc,Rattazzi:2008pe,Rychkov:2009ij,Poland:2011ey,Beem:2013qxa,Chester2014,Kos2014,Kos2014a}).
The basic idea is to use crossing symmetry constraints to derive restrictions on the spectra of hypothetical conformal field theories. Since only the basic properties of unitarity, crossing and conformal symmetry are used, this program has been able to find highly non-trivial {\em non-perturbative} information on a variety of fixed points. 

The vast majority of results has so far involved  numerical calculations on a computer. Depending on the degree of truncation of the problem and the accuracy desired in the results, the computational power required can vary from a single laptop \cite{Rattazzi:2008pe} to hundreds of cores \cite{El-Showk2014a}. Software used so far has included Wolfram's Mathematica, IBM's CPLEX, SDPA-GMP and implementations of the simplex algorithm in C++ and Python. The present note introduces {\tt JuliBootS}, an open-source, {\tt Julia}-based package for performing numerical bootstrap computations. One goal is to make these methods widely available, and as simple to use as possible, so that the interested reader may jump into this exciting research program straight away without too much worry about what's going on under the hood. However, another purpose is to have a common framework for developing these methods publicly, so that higher computational efficiency, precision and scope may be reached in the near future. {\tt Julia} is a fast-developing language which aims to combine the ease of use of Python with the speed of C. We chose it for its elegant multiple-dispatch features, calling of both Python and C code, and quite importantly for the end user, the fact that use and installation is extremely simple: there is no compilation of code required at any point, and as such after {\tt Julia} is on the system {\tt JuliBootS} will run out of the box.

Here's a quick layout of this note: in the section \ref{intro} we make a pedagogical introduction to the bootstrap program -- what the problem is and how we can solve it. The short section \ref{general} describes how to get {\tt JuliBootS} and install it as well as the general layout of the code. The following sections describe several pieces of the code -- conformal blocks, the linear solver and the minimization algorithm. Section \ref{bs} is by far the most important, discussing a number of applications of the code to real-world examples.

\subsection*{Quickstart guide!}

If the reader wants to immediately get his hands dirty, we recommend she skips to section \ref{install} for installation instructions, and then to section \ref{bs} to start bootstrapping right away. Sample conformal block tables are provided with the package, and extra ones can be generated with the {\em Mathematica} notebook {\tt ComputeTables.nb} that comes with {\tt JuliBootS}.

The {\tt JuliBootS} package is still being developed and as such there is large room for improvement -- suggestions, comments and bug reports are all greatly appreciated. If you wish to contribute do not hesitate to contact me, or submit a request on GitHub at:
\bea
\mbox{{\tt https://github.com/mfpaulos/JuliBoots}.}
\eea
Happy bootstrapping!

\section{A pedagogical introduction to the numerical bootstrap}
\label{intro}

\subsection*{Crossing symmetry}
We consider the four point function of a scalar operator of dimension $\Delta_\sigma$ in a conformal field theory (CFT), 
\bea
\langle \sigma(x_1)\sigma(x_2)\sigma(x_3)\sigma(x_4)\rangle=\left(\frac{x_{13}^2\,x_{24}^2}{x_{12}^2 \, x_{23}^2\, x_{34}^2\,x_{41}^2}\right)^{\Delta_\sigma}\, f(u,v).
\eea
The undetermined function $f(u,v)$ is forced by symmetry to depend only on conformally invariant cross-ratios, here taken to be
\bea
u=\frac{x_{12}^2\,x_{34}^2}{x_{13}^2\,x_{24}^2}\,,
\qquad
v=\frac{x_{14}^2\,x_{23}^2}{x_{13}^2\,x_{24}^2}.
\eea
with $x_{ij}\equiv x_i-x_j$. The existence of a convergent operator product expansion (OPE), leads to a stronger statement: we can decompose the function in terms of conformal blocks, which capture contributions of individual primaries\footnote{Primaries are highest weights of the conformal group $SO(d+1,1)$ -- usually called quasi-primaries in $d=2$.} $\mathcal O$ and their descendants in the $\sigma \times \sigma$ OPE. Since the OPE can be taken in different ways we get inequivalent decompositions which must nevertheless match -- a strong constraint. For instance, taking the OPE in the (12) and (14) channels leads to:
\bea
f(u,v)=\sum_{\Delta,L} \lambda^2_{\Delta,L}\, v^{\ds} G_{\Delta,L}(u,v)  =\sum_{\Delta,L} \lambda^2_{\Delta,L}  u^{\ds} G_{\Delta,L}(v,u) 
\eea
In this expression, we sum over contributions of primaries with conformal dimension $\Delta$ and spin\footnote{Two identical scalars can only couple to operators in traceless symmetric representations of the rotation group, which are uniquely labeled by their spin.} $L$. The functions $G_{\Delta,L}(u,v)$ are the conformal blocks themselves, which are in principle completely determined by conformal symmetry although technically they are not so easy to compute. Finally the coefficients $\lambda_{\Delta,L}$ are the OPE coefficients appearing in three-point functions of the form $\langle \phi \phi \mathcal O_{\Delta,L}\rangle$. These, together with the spectrum of operators, are the dynamical data which characterizes a given theory.

The idea now is to take a step back and think of these equations not as being associated to a particular theory, but rather as a set of constraints which must hold generically. Equivalently, they give us sum rules which must hold for any consistent CFT spectrum. Our strategy will be to use these equations as a starting point and understand what they imply about generic CFTs. At first sight this might seem hopeless: there is a continuously infinite set of constraints for a continuously infinite set of parameters; the equations involve the conformal blocks which can be determined analytically only in certain cases; and finally, there is no expansion parameter in sight. Remarkably, all these obstacles can be overcome.

\subsection*{Linear Programs}
We begin by rewriting the constraints slightly:
\bea
\sum_{\Delta,L} a_{\Delta,L}\, F^{\ds}_{\Delta,L}(u,v)=0.\label{linear}
\eea
Here we have defined
\bea
a_{\Delta,L}\equiv \lambda_{\Delta,L}^2,\qquad F^{\ds}_{\Delta,L}(u,v)\equiv \, v^{\ds}\,G_{\Delta,L}(u,v)-u^{\Delta_\sigma}\,G_{\Delta,L}(v,u).
\eea
Equation \reef{linear} is a {\em linear} equation for the $a_{\Delta,L}$. One solution of the constraints would be to set all coefficients to zero. However, this is not allowed, since the identity operator is always present in the OPE with unit coefficient -- $a_{0,0}=1$. Also, although the equations are linear we are still faced with an infinite number of parameters and constraints. But suppose we truncate the equations somehow -- for instance by doing a Taylor expansion to some fixed order $N$ -- then as long as we have $N$ linearly independent ``vectors'' given by a set of the truncated $F^{\ds}_{\Delta,L}$ we are done -- we can solve the linear system. However, this is not really the end of the story. Very generically, and in fact overwhelmingly so, doing this will lead to a solution with at least some negative $a_{\Delta,L}$. This is a problem, since these are squares of OPE coefficients. In particular, if we now make the crucial assumption of {\em unitarity}, all $\lambda_{\Delta,L}$ can be chosen real, which in turn implies positivity of the $a_{\Delta,L}$. 

Unitarity then is what makes the problem interesting. It allows us to extract useful information from the sum rule by imposing {\em positivity} of the coefficients. The first step of any bootstrap calculation is to rewrite the constraint equations in a way which makes positivity manifest. In cases where one is dealing with a single correlation function, then the analysis above suffices. When fields are charged under a global symmetry we have to work a bit harder~\cite{Rattazzi:2010yc}, but it is still possible. For multiple correlators it is even harder, but one can still do it -- see \cite{Kos2014a} for very interesting developments in this area\footnote{In this case the positivity constraints can be phrased in terms of positive matrices, which leads to a class of problems known as semidefinite programming.  The {\tt JuliBootS} package is designed to focus exclusively on linear programs.}.

Going back to our single correlation function example, positivity implies that the actual problem that we need to solve takes the schematic form
\bea
\sum_{i} a_i \mbf v_i=0, \qquad \forall_{i\neq 0}\, a_i\geq 0,\qquad a_0=1. \label{lp}
\eea
A few comments are in order. Firstly, here the vectors $\mbf v_i$ are the $N$-dimensional truncations of the functions $F^{\ds}_{\Delta,L}(u,v)$ -- typically $N$ different derivatives evaluated at fixed $u,v$. Other choices of basis are possible, although numerically this approach turns out to be particularly convenient. Secondly, in the summation we should allow all possible $\Delta$ and $L$ consistent with unitarity. In particular the allowed sets $\Delta(L)$ are continuous and unbounded from above, so the above should really be a series (for spin $L$) of integrals (for $\Delta$). They are however bounded from below by the constraint that we should only allow unitary representation of the conformal group. This imposes
\bea
\Delta\geq \left\{ 
\begin{tabular}{c} 
$(d-2)/2, \qquad  L=0$\\
$L+d-2, \qquad L> 0.$
\end{tabular}
\right. \label{unitbounds}
\eea

The problem above is a special type of  {\em linear program}: minimization of a linear cost function subject to linear equations and inequalities. In this case we have a trivial cost function -- we are only interested in determining if the set of constraints has a feasible solution. This is what we must do for instance in computing bounds on operator dimensions. However, we will also be interested in bounding the values of (squares of) OPE coefficients. In this case we get a more general linear program:%
\bea
\mbox{maximize}\, a_k\quad \mbox{s.t.} \quad \sum_{i} a_i \mbf v_i=0, \qquad \forall_{i\neq 0}\, a_i\geq 0,\qquad a_0=1. \label{opelp}
\eea
In any case, these kinds of problems have been well studied since the 1930's, since they appear in such practical settings as determining how to optimize the production and profit of a factory. This means that there are efficient methods for solving them, for instance Dantzig's simplex method\footnote{For an introduction and references, go no further than the Wikipedia page, {\tt http://en.wikipedia.org/wiki/Simplex\_algorithm.html}.}, which is the method we will explain and use in later sections.  

To summarize, we have reduced the problem of solving the crossing symmetry constraints to a linear program which can in principle be solved numerically. However, before we solve the linear program, we need to actually determine the ingredients that make it up -- the conformal blocks, which in turn determine the $F_{\Delta,L}^{\ds}$ functions.

\subsection*{Conformal blocks}

The computation of conformal blocks has seen major progress in recent years, especially for ``vanilla'' blocks, with traceless symmetric spin and without supersymmetry -- \cite{Costa:2011dw,Dolan2011,DO2,DO1,Hogervorst:2013kva,Hogervorst:2013sma,Osborn2012,Simmons-Duffin2014}. In particular it is possible to derive efficient recursion relations for derivatives of conformal blocks at fixed $u=v$ for any spacetime dimension~\cite{ElShowk:2012ht}. Reference \cite{Kos2014} found a very efficient recursion relation for blocks inspired by a similar one in two-dimensions \cite{Zamolodchikov1987,Zamolodchikov1984}, whereas \cite{Hogervorst:2013kva,Hogervorst:2013sma} uses the fact that conformal blocks are eigenvectors of the conformal casimirs to derive similar relations. The upshot is that there are now methods for deriving representation of the conformal blocks and their derivatives that take the form:
\bea
\partial_u^m \partial_v^n G_{\Delta,L}(u_0,v_0)=u_0^{\frac{\Delta}2}\, \left(\sum_k^{P} a^{(k)}_{mn}\Delta^k+\sum_j \frac{b^{(j)}_{mn}}{\Delta-\Delta_j}\right).
\eea
That is, (non-supersymmetric) conformal blocks for traceless symmetric spin representations are well understood, in the sense that we have good control over their values and their derivatives at fixed $u,v$, by representing them in terms of rational functions in $\Delta$. It is this representation that is currently used by \jb. An important remark is that these results can be derived for any dimension -- even fractional ones.

\subsection*{Truncations}

Now let us turn to solving the equations themselves. Here we are faced with the problem that the full set of ``vectors'' (conformal blocks) is continuously infinite and unbounded. The vectors themselves are functions, and hence have a continuous number of degrees of freedom. The first thing to do is, as explained, to consider a finite truncation of the problem -- usually by working with the Taylor expansion of the $F$ functions. Since there are efficient methods for computing derivatives of blocks, we are done. Next, we must truncate the spectrum of allowed operators. After imposing the unitarity bounds \ref{unitbounds}, we still have arbitrarily large conformal dimensions and spins in the sum rule. This is a bit of an issue, at least for the simplex method which requires a bounded search space. We solve this problem by imposing cut-offs in both spin and conformal dimension\footnote{The truncation in conformal dimensions is not required if solving the equations using semi-definite programming methods \cite{Poland:2011ey}, but here we will concentrate on Linear Programming.}. 

What is the effect of this truncation? Suppose one attempts to solve the equations and does not find a solution. Then we can't be really sure whether this is a consequence of the constraints or of the cut-offs: perhaps allowing more vectors would allow one to find a solution. However there are ways of giving us confidence in this approximation. One basic check is to verify that the results do not change by moving the cut-off. One may also in some cases include the asymptotic form of the conformal blocks for large $\Delta$ and/or spin. In practice one finds that relatively small cut-offs of the order of a forty or so are sufficient. This is to be expected, since we are always considering truncations of the problem down to some finite number $N$ of equations, and so we may hope that higher spin and higher dimension operators shouldn't play much of a role in the problem, at least if we choose these equations wisely. In actual applications this is indeed the case so that truncation is not really an issue. This is not an accident, as the OPE has been shown to be convergent exponentially fast, at least away from the light-cone \cite{Pappadopulo:2012jk}.

Finally, we must deal with the fact that we have a continuous (and therefore infinite) search space. In past work, this was mostly handled by discretising, considering conformal blocks on some finite, but large grid. There are two ways around this. Firstly, the use of semi-definite programming methods do away with this restriction entirely as we can work directly with certain polynomial representations for the conformal blocks  \cite{Poland:2011ey}; and secondly, it is in fact possible to adapt the basic simplex algorithm to allow for continuous search spaces, as shall be described in a later section \cite{El-Showk2014a}. 

\subsection*{Bootstrapping}

Now that all the ingredients are in place, we are finally ready to get results. We start with the bootstrap bread-and-butter: deriving bounds on operator dimensions. The basic idea is to impose constraints on the vectors allowed in the linear program \ref{lp} until a solution can no longer be found. As an example, we may set a gap on the first scalar -- we exclude from the sum rule all spin-0 operators with dimension smaller than some variable $\Delta_{gap}$. Increasing the gap one eventually finds that it is not possible to solve for the sum rule. Now, what does this mean? Well, suppose our gap is such that the sum rule can be solved. Then we haven't learned much: since we are always working with a truncation of the problem to a finite number of constraints, it is possible that the addition of more constraints may eliminate this solution. However, if we {\em cannot} find a solution then we can make a definite statement, since augmenting our truncated equations can only make the problem more restrictive. In this way it follows that for any given $N$ we can find a perfectly valid upper bound, a bound which can only decrease as $N$ is increased. Finally, by varying the dimension of the external scalar $\Delta_\sigma$ we get a curve $(\Delta_\sigma, \Delta_\epsilon)$ which often can show interesting features such as kinks.

An even simpler variation is to bound OPE coefficients. Here we simply solve the linear program \ref{opelp} directly. Similarly to the previous case, here one gets again a valid upper bound on the OPE coefficient (squared) for each value of $N$. Let us comment that a particularly interesting case for OPE maximization is that of the stress tensor. This not only restricts us automatically to the interesting (but not exhaustive) set of {\em local} CFTs, but also gives us a lower bound on the central charge of CFTs.

We should mention that it may be interesting in either case to impose extra constraints before deriving the bounds, by imposing gaps or cuts in the sets of vectors allowed. This can be done for instance if we have some information about the spectrum of the theory or class of theories we are interested in, leading to more stringent results. 

A very important fact is that once the linear programs are solved we have found a proper solution to the crossing symmetry constraints: a set of blocks and OPE coefficients which add up to the identity vector. In particular this set gives some approximation to a conformally invariant correlation function. When bounding operator dimensions, this solution is generically not unique -- for a fixed gap there are in principle an infinite set of solutions to crossing. The exception is when we tune the gap such that we are sitting precisely at the boundary of the allowed region \cite{ElShowk:2012hu}, where the solution does become unique. This is very interesting, for if we know that some actual CFT is supposed to be sitting at that point we can reconstruct its spectrum. Excitingly this seems to be the case for the critical Ising model and several other examples. The analysis of the spectrum as the dimension of the external scalar field $\Delta_\sigma$ is varied can also give some insight into features of the bounds such as the kinks mentioned above. 

\section{Generalities}
\label{general}
\subsection{Installation}
\label{install}

The \jb \ package has been developed in the {\tt Julia} language, and so in order to run the former the latter must be installed. The latest version of the {\tt Julia} language can be downloaded from the official website {\tt http://julialang.org}. As of the date of this note, the current release is v0.3.0 and can be obtained from {\tt http://julialang.org/downloads}. Installation instructions are given on the website and should be straightforward on most platforms. Basic knowledge of the language is convenient but not absolutely necessary for most applications of \jb. We strongly advise the excellent manual available at {\tt http://docs.julialang.org/en/release-0.3/} 

During development we have found {\em Julia Studio} to be a convenient open-source IDE. It is available at {\tt http://forio.com/labs/julia-studio}, and comes bundled with {\tt Julia} v0.3.0. The page {\tt http://forio.com/labs/julia-studio/tutorials} has useful tutorials for {\tt Julia}.

As for the {\tt JuliBootS} package itself, it can be downloaded from its GitHub repository at {\tt https://github.com/mfpaulos/JuliBoots}. The current release is version 1.0. If {\tt Julia} is installed, there's nothing left to do after download -- the package runs out of the box with no compilation required. After running {\tt Julia} on the \jb directory, the package can be loaded with the command:
\bec
julia> using juliboots
\eec
In section \ref{bs} we have compiled a set of examples which should allow the reader to get a good handle on how to use the package. We hope to provide a more detailed reference guide in the near future.

\subsection{Global structure}

\begin{figure}
	\centering
		\includegraphics[width=15 cm]{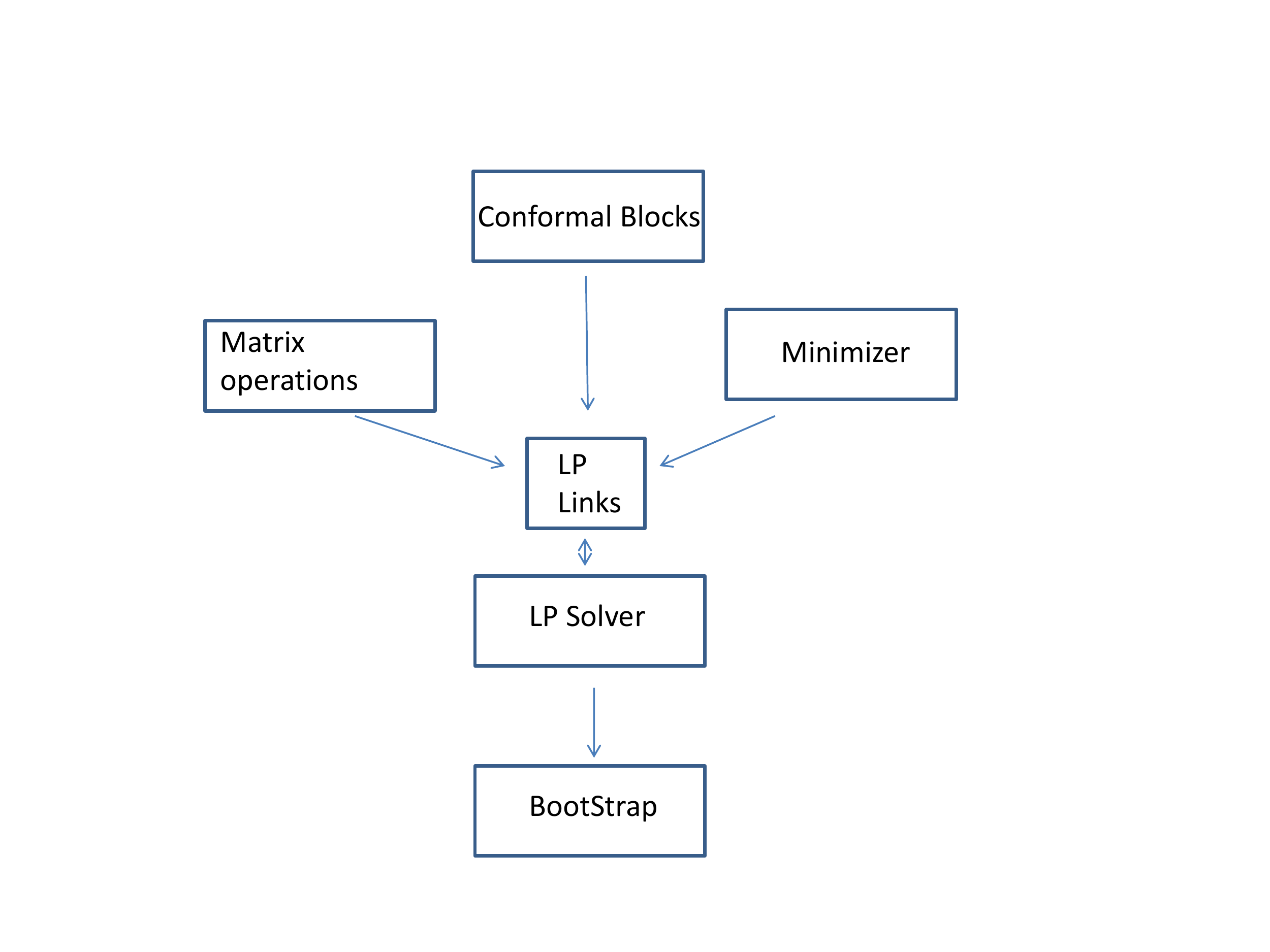}
	\caption{Global code structure.}
	\label{fig:codestructure}
\end{figure}

The {\tt JuliBootS} package has been designed with modularity in mind. The code is composed of several independent modules which can be turned on or off at will -- so that if in the future one implements a better matrix inversion algorithm, a different conformal block representation or even a different linear program solver, it should be fairly straightforward to do the necessary changes. The structure of the code at this point in time is outlined in figure \ref{fig:codestructure}.

Currently, the main engine driving the code is a simplex method solver capable of handling discrete and continuous search spaces. This is included in the LP solver module. The solver communicates with other modules via a sub module LP Links. The solver needs to know about how conformal blocks are encoded (what {\tt Julia} Types it should allow); how to minimize one-parameter functions (as part of the simplex algorithm); and how to perform various kind of matrix operations, such as inversion and LDU decomposition. Finally, there are various higher level functions defined in the Bootstrap module which actually allow the end user to set up and solve various kinds of problems -- such as obtaining bounds, maximizing OPE coefficients, etc. In the next few sections we will describe in some detail some of these modules.

\section{Conformal blocks}
\label{cb}
We start off with the basic ingredients of any conformal bootstrap calculation -- the conformal blocks themselves. Knowledge of the conformal blocks is required to form the vectors that will eventually be fed into the Linear Program solver. Although the solver is agnostic to the kinds of vectors we use, currently we employ the strategy of Taylor expanding the conformal blocks around the symmetric point $u=v=1/4$, so that the vectors are made up of various kinds of derivatives.

\subsection{Rational representation}

As outlined in section \ref{intro}, previous work has given us ways of computing the derivatives of conformal blocks efficiently. The method we use has been described in \cite{Hogervorst:2013kva,Hogervorst:2013sma} and implemented in \cite{El-Showk2014a}. The basic idea is to use the fact that conformal blocks are eigenfunctions of the Casimir operators. Defining $u=z \bar z, v=(1-z)(1-\bar z)$, and 
\bea
z=\frac{a+\sqrt{b}}2, \qquad \bar z=\frac{a-\sqrt{b}}2,
\eea
then the Casimir equations imply recursion relations which allow us to determine the block and its $a$ derivatives as a power series in $a$ for $b=0$. Actually, there is a more efficient way to do this procedure where one partially resums the set of $a$ derivatives by working with a new variable
\bea
\rho=\frac{2-z-2\sqrt{1-z}}z.
\eea
We then use another Casimir-derived recursion relation which gives $b$ derivatives in terms of $a$ derivatives for fixed $a$. Concretely, the blocks take the form:
\bea
\partial_a^m \partial_b^n G_{\Delta,L}(a=1,b=0)=(4\rho)^\Delta\, \left(\sum_k^{P} a^{(k)}_{mn}\Delta^k+\sum_j \frac{b^{(j)}_{mn}}{\Delta-\Delta_j}\right),\label{derg}
\eea
where $\rho$ should be evaluated at the symmetric point, $\rho(z=1/2)=3-2\sqrt{2}$. In our conventions, the conformal blocks are normalized such that for $z=\bar z\ll 1$ we have
\bea
G_{\Delta,L}\simeq (z \bar z)^{\Delta}(1+\ldots).
\eea
In particular this does not match the conventions of Dolan and Osborn \cite{Dolan2011,DO2,DO1}, but it is simple to go from one to the other
\bea
G_{\Delta,L}=G_{\Delta,L}^{DO} \times \frac{\Gamma(d-2+L)\Gamma\left(\frac{d-2}2\right)}{2^L\,\Gamma(d-2)\Gamma\left(\frac{d-2+2L}2\right)}.
\eea
The recursion relations are implemented in a {\tt  Mathematica} notebook\footnote{We hope that in future versions there will be a direct {\tt Julia} implementation of this computation so as to avoid the use of commercial software.} -- {\tt ComputeTables.nb} -- 
 which is included with \jb. The notebook itself has instructions on how to use it. The upshot is that after selecting a maximum number of $m,n$ derivatives (\mmax,\nmax), a maximum number of spins, and the spacetime dimension, the output of this notebook is a table file which can be loaded by \jb. Because of the way in which the recursion relations work, the derivatives one obtains take a triangular form:
\bea
\{(m,n): \quad m=0,\ldots,2(n_{\mbox{\tiny max}}-n)+m_{\mbox{\tiny max}}; \qquad n=0,\ldots, n_{\mbox{\tiny max}}\}
\eea
In particular setting \mmax=0 doesn't mean one will not compute $m$ derivatives.

Finally, notice that given the derivatives of the blocks, it is possible to obtain those of the $F_{\Delta,L}^{\Delta_\sigma}(z,\bar z)$ functions by a simple linear operation. We call this operation convolution, since we convolve the $G$ vectors with a kernel depending on $\Delta_\sigma$ to obtain the $F$ vectors. So, the key point is computing the derivatives of the blocks themselves, which is done in the {\em Mathematica} notebook.. The convolution operation however is implemented in \jb.

Let us try all this out. We begin by starting the code,
\bec
julia> using juliboots
\eec
and now we load up a simple table:
\bec
julia> tab=loadTable("./Tables/D3\_n3m1\_L20.txt")\\
CB derivatives table - D = 3.0\\
\ \\
Convolved with sigma = NaN\\
(nmax,mmax) = (3,1)	 20 components\\
Lmax = 20	 Odd Spins - false\\
Precision: 265
\eec
As we can see the output is a derivatives table object. We can examine which kinds of $(m,n)$ derivatives it contains with
\bec
julia> tab.ders\\
20-element Array{(Int64,Int64),1}:\\
 (0,0)\\
 (1,0)\\
\ \ $\vdots$\\
 (1,3)
\eec
The table itself is contained in $\tt tab.table$. It is an array of conformal block vector functions in the rational representation, one element per spin. For instance, to evaluate the derivatives above for the spin-0 conformal block at some particular value of conformal dimension $\Delta$, we may do
\bec
julia> spin0=tab.table[1];\\
julia> value(spin0,BigFloat(1.2))
\eec
This gives the derivatives at $\Delta=1.2$. More precisely, the values correspond to the derivatives given in \ref{derg}, multiplied by an extra factor $2^{m+2n}$ for convenience. We can dig deeper into the entrails of the conformal block to find its rational function representation. Picking a particular component,
\bec
julia> component=spin0[(1,0)]\\
CB\_Q - Label = "Vanilla N=0" - 	Spin = 0	 (m,n) = (1,0)\\
julia> rational=component.func
\eec
The first type is {\tt CB\_Q} which stands for a conformal block in a rational (Q) representation. Inside such objects are contained the rational functions themselves, which are of type {\tt QFunc}. This in turn is made up of a polynomial and a series of poles -- {\tt rational.poly} and {\tt rational.poles} -- all of which can also be evaluated at specific values of their arguments using method $\tt value(obj, arg)$.

\subsection{Convolution}

Before we pass on our conformal blocks to the linear solver, we must {\em convolve} them first. Convolution is the name usually given to the operation: 
\bea
G_{\Delta,L}(u,v) \to v^{\Delta_\sigma} G_{\Delta,L}(u,v)\pm u^{\Delta_\sigma} G_{\Delta,L}(v,u) 
\eea
This introduces the dependence on the external operator dimension $\Delta_\sigma$. In {\tt JuliBootS} we can convolve a whole table directly at a specific value of $\Delta_\sigma$. We must specify a sign -- the plus sign is required for instance when bootstrapping correlators with global symmetry. The command is:
\bec
julia> sigma=BigFloat(0.518);\\
julia> convolved=convTable(sigma,tab,-1)\\
CB derivatives table - D = 3.0\\
\ \\
Convolved with sigma = 0.518\\
(nmax,mmax) = (3,1)	 10 components\\
Lmax = 20	 Odd Spins - false\\
Precision: 265
\eec
Notice how the table has only 10 components now. This is because all even $m$-derivatives vanish by symmetry. Had we chosen the opposite sign, it would have been the odd $m$-derivatives which would be automatically zero. If required, we can also convolve a single vector of conformal blocks (i.e. for a particular spin)
\bec
julia> convolved\_spin0=convBlock(sigma,spin0,-1)
\eec
The output is a convolved conformal block vector of derivatives, which has its own type $\tt convVec\_Q$. It is almost the same as the unconvolved type {\tt CBVec\_Q}, but it carries an extra piece of information to tell us which value of $\Delta_\sigma$ it corresponds to. We can also evaluate it at a value of $\Delta$ using method $\tt value$, as mentioned before.

\section{The solver}
\label{solver}

In this section we will discuss the main engine of the code: its simplex method Linear Program solver. 
The solver is based around an important type -- {\tt LinearProgram} -- and an important method -- {\tt iterate!}. The first contains all the information regarding a particular linear program that we wish to solve; while the second systematically solves it by reducing the cost. The solver is currently set up for arbitrary precision using {\tt BigFloat} numbers, but we hope to implement a floating-point version in the near future. The simplex method solver is an independent implementation with some variations of the algorithm first described in \cite{El-Showk2014a}, which has been already implemented in Python and C++.

\subsection{Linear Programs}
Linear programs are encoded in simplex method based {\tt LinearProgram} types. Since this is such an important type, let us describe its contents in detail:
\begin{itemize}
\item[] Type {\tt LinearProgram}
\begin{itemize}
\item {\tt lpFunctions} - an array of vector functions.
\item {\tt lpVectors} - an array of discrete vectors.
\item {\tt target} - an $n$-dimensional vector $\mbf t$. 
\item{\tt solVecs} - a set of $n$-dimensional vectors $\mbf v_i^*$ such that $\sum a_i \mbf v_i^*=\mbf t$ with $a_i>0$.
\item{\tt coeffs} - the coefficients $a_i$.
\item{\tt invA} - the inverse of the matrix $A\equiv \left( \mbf v_1^*| \ldots | \bf v_n^*\right)$
\item{\tt functional} - the current linear functional as given in equation (\ref{linfunc}) below.
\item{\tt label} - a description for this linear program.
\end{itemize}
\end{itemize}
A linear program is the following problem:%
\bea
\mbox{min} (C\equiv \sum_{i\in S} c_i a_i) \quad \mbox {s.t.} \quad 
\sum_{i\in S} a_i \mbf v_i=\mbf t, \qquad a_i>0
\eea
Here $S$ is a {\em search space}, a set of vectors which we allow in the sum rule. We will be considering search spaces made up both of discrete and continuous sets of vectors, so that $i$ can be a continuous label. The first two entries in {\tt LinearProgram} comprise the search space. In particular the first entry is an array of vector functions. Each vector function encodes a set of (convolved) conformal block derivatives varying over some range in conformal dimension, for a fixed spin. Each vector in the search space has an associated cost, which is used in the computation of the total cost $C$. Similarly vector functions also have associated cost functions.

To solve such a problem we need first to find a feasible solution, i.e. a specific set of $n$ vectors $\mbf v_i^*$ such that the sum rule is satisfied. As such, before attempting to minimize the cost function we need to solve an auxiliary problem:
\bea
\mbox{min} (C'\equiv \sum_{j\in \mbox{ aux}} c_j b_j) \quad \mbox{s.t.} \quad
\sum_{j\in \mbox{aux}} b_j \mbf w_j+\sum_{i\in S} a_i \mbf v_i=\mbf t, \qquad a_i,b_i>0
\eea
The auxiliary set {\em aux} is comprised of $n$ vectors $\mbf w_j$ such that in components $w_{j,i}=sgn(t_i) \delta_{i,j}$. The cost vector $\tilde c_i$ is irrelevant as long as it has all positive components. Typically we will set it to the vector with all unit entries. In this way, it is clear that a feasible set is $\mbf v_i^*\equiv \mbf w_i $, since the constraints are satisfied with $\forall_i a_i=0, b_i=|t_i|$. The simplex algorithm now proceeds by altering the set $\mbf v_i^*$, systematically reducing the cost. If we can reduce the cost $C'$ in this auxiliary problem to zero, then we have found a solution to the sum rule which does not involve any auxiliary vectors. We can then proceed to switch on the costs $c_i$ for all the $\mbf v_i$ and solve the original problem in the exact same fashion.

In many cases however we are simply interested in checking if there is at least one solution to the set of constraints. This is what happens if one only wants to check that the crossing symmetry constraints can be solved given some set of allowed vectors in the sum rule. In this case it is sufficient to solve the auxiliary problem. The second step is required in case we wish to maximize an OPE coefficient, or some combination of OPE coefficients.

\subsection{The Simplex Method}
\label{simplex}
We shall now describe the simplex method in some detail, using the code as our guide. This will help us better understand how the code is structured. For convenience, we have also summarized the method in some detail at the end of this section. Our goal will be to find a solution to crossing symmetry for the case of a correlator of four identical scalars,
\bea
\sum_{\Delta,L} a_{\Delta,L}\, F^{\ds}_{\Delta,L}(u,v)=-F^{\ds}_{0,0}(u,v),\label{basicprob}
\eea
where the right-hand side represents the contribution of the identity operator. 

As a first step we need to initialize a {\tt LinearProgram} type. This means loading a table of conformal blocks, convolving them to obtain the vectors of $F$ derivatives, but also setting up the auxiliary vectors $\mbf w_j$. All this can be done with the method $\tt setupLP$:
\bec
julia> using juliboots\\
julia> sigma=BigFloat(0.518);\\
julia> prob=setupLP(sigma,"./Tables/D3\_n3m1\_L20.txt")\\
Setting up LP...\\
Done\\
Linear Program:Basic Bound\\
D = 3.0	sigma=0.518	(m,n) = (1,3)	Lmax = 20	Odd spins: false
\eec
We can examine the contents of the search space using $\tt prob.lpFunctions$ and $\tt prob.lpVectors$. The first shows us there are functions defined for each even spin up to $20$, with ranges varying from the unitarity bounds up to some cut-off, which is set to 70 by default\footnote{Defaults can be changed in the {\tt consts.jl} file.}; the second shows us a set of 10 auxiliary vectors, the same as the number of components of each vector. These are currently the vectors that make up the feasible solution, as can be checked with {\tt prob.solVecs}. A convenient method is {\tt status}, which tells us about the current vectors in the solution and which coefficients they appear,
\bec
julia> status(prob)
\eec
One can check that the coefficients $\tt prob.coeffs$ are precisely the same as the absolute values of the components of the target given by $\tt prob.target$. That the target is (minus) the identity vector can be checked by evaluating the spin-0 $\tt lpFunction$ at $\Delta=0$ and comparing:
\bec
julia> id=value(prob.lpFunctions[1],0.);\\
julia> id+prob.target\\
10-element Array\{BigFloat,1\}:\\
 0e+00\\
 0e+00\\
 \ \ $\vdots$\\
 0e+00
\eec
The cost of an auxiliary vector is set to unity by default. Hence the total cost of the linear program should be the sum of the coefficients:
\bec
julia> cost(prob)-sum([abs(x) for x in prob.target])\\
0e+00 with 212 bits of precision
\eec
In order to lower the cost $C$, we try to bring in one of the non-auxiliary vectors into the solution, by switching on some $a_k$. But we must do this while satisfying the constraints:
\bea
&&\sum_{i=1}^n b_i \mbf v_i^* + a_k \mbf v_k=\mbf t \Leftrightarrow b_i= (A^{-1} \mbf t)_i-a_k (A^{-1} \mbf v_k)_i \label{bj}\\
&\Rightarrow & C=c_0+a_k \Phi \cdot \mbf v_k
\eea
with $c_0$ the original cost, and $\Phi$ the {\em linear functional}:
\bea
\Phi_k=-\sum c_i A^{-1}_{i k}.\label{linfunc}
\eea
Here $A^{-1}$ is the inverse of the matrix whose columns are the vectors making up the current solution to the constraints, and the $c_i$ their respective cost. In our current setup we have set the costs of all vectors except the auxiliary ones to zero, and hence the functional is non-zero as long as there is an auxiliary vector in the current solution. Since $a_k$ must be positive, the previous equation shows that if we want to decrease the overall cost we must switch on a vector such that the functional acting on it is negative. This is then the first step in the simplex method: to find such a vector. If at any point we cannot succeed in doing this, the method stops. There are then two possibilities. The first is that we have reached a point where all vectors in the current solution are physical, since then the functional is trivially zero. In the second case we have shown that:
\bea
\Phi\cdot \mbf t<0, \qquad \forall_k \Phi\cdot v_k>0  \quad \Rightarrow \not \exists \{a_i>0\} : \sum a_i \mbf v_i=\mbf t
\eea
i.e. there is no solution to the problem. Hence we see that the functional is capturing very important information.

Let us go back to our {\tt JuliBootS} example. We want to see if there is at least one vector such that the functional acting on it is negative. Typically there are many of those, and we must choose one of them. One way of choosing is to pick one for which the variation of the cost will be the fastest as we turn on its coefficient -- the vector with {\em minimum reduced cost} (MRC), which is simply the value of the functional acting on the vector. The method for doing this is:
\bec
julia> invec=findMRC(prob)\\
(LPVector - (5.0...e-01 with 212 bits of precision,"L=0"), cost - 0e+00
,-1.498...e+30 with 212 bits of precision)
\eec
Its output is a doublet, made up of an {\tt LPVector} object, and its corresponding reduced cost, which is negative. So we see that at this stage, we haven't ruled out the problem as insoluble. This is good, since as we have not imposed any constraints at all on the set of allowed operators, we know that at least the generalized free field CFT should be allowed.

The simplex method now proceeds by determining which of the vectors in the solution should be swapped with the newcomer. This is simple to determine: we merely increase $a_k$ until some critical value $x$ such that one of the coefficients $b_j$ vanishes, according to equation (\ref{bj}); that is:
\bea
x=\mbox{max}_{a_k}: \forall_i b_i= (A^{-1} \mbf t)_i-a_k (A^{-1} \mbf v_k)_i\geq 0 \label{xval}
\eea
Since all coefficients must be positive, we must stop at this point. We have then succeeded in bringing in a new vector into the current solution, and reducing the cost by an amount $x \times \mbox{MRC}$. In {\tt JuliBootS} we can determine which vector will exit the solution via
\bec
julia> outvec=findBVar(prob,invec[1])\\
(1.3326\ldots43e-30 with 212 bits of precision,10)
\eec
The output is again a doublet, made up of a real number and an integer. The integer tells us which vector in the current solution will exit. In this example, the vector is the tenth out of the current ten vectors making up the solution, so that the first $b_i$ to reach zero when increasing $a_k$ is $b_{10}$. The real number tells us the value of $a_k$ for which this happens, i.e. $x$ in equation (\ref{xval}). At this point we have new vectors making up the solution, and as such we should update the linear functional according to equation (\ref{linfunc}). At this point we are exactly as at the beginning of the process, but with a smaller overall cost. We can repeat the procedure until the cost no longer varies.

This whole procedure is encoded in a single method, ${\tt iterate!}$. It takes a Linear Program object and performs the simplex method on it $m$ times, updating the problem in the process (conventionally methods with ! in the name change their modify their arguments). If we want to save the original problem, we'd better make a backup copy first:
\bec
julia> backup=mcopy(prob);\\
julia> iterate!(prob,1,method="mrc");
\eec
If we now check the {\tt status} of {\tt prob} we can see precisely that the 10th vector in the solution has been replaced by the vector we found previously with {\tt findMRC}. We can now keep going until a complete solution is found
\bec
julia> iterate!(prob,500)
\eec
After a few iterations the minimum possible cost is achieved. In this case we can check it is zero, and that indeed we have found a solution containing no auxiliary vectors, once again using method {\tt status}. A more accurate list of operators and their coefficients can be obtained from {\tt prob.solVecs} and {\tt prob.coeffs}, or all at once with
\bec
julia> solution(prob)
\eec

\subsection*{Summary}
To solve the problem:
\bea
\mbox{min} (C\equiv \sum_{i\in S} c_i a_i) \quad \mbox {s.t.} \quad 
\sum_{i\in S} a_i \mbf v_i=\mbf t, \qquad a_i>0
\eea
we apply the following sequence of steps:
\begin{itemize}
\item At any given point we have a set of $n$ vectors $\mbf v_i*$, whose columns form a matrix $A$. These vectors make up a solution to the constraints with some coefficients,
\bea
\sum_{i=1}^n a_i \mbf v_i^*=t \Leftrightarrow A\cdot \mbf a=\mbf t
\eea
 The total cost is the product of these coefficients with the cost of each vector $C=\sum_{i=1}^n c_i^* a_i=\mbf c^*\cdot A^{-1}\cdot \mbf t$.
\item We attempt to bring in a vector from the search space $S$ into the basis such that the cost is reduced. This is possible if for some vector $\mbf v_k \in S$ we have negative reduced cost:
\bea
RC=c_k+\Phi\cdot \mbf v_k<0 \label{negrc}
\eea
with $c_k$ the cost associated to the vector, and $\Phi$ the linear functional given by $$\Phi\equiv -\mbf c^*\cdot A^{-1}.$$
\item We turn on the coefficient $a_k$ of one such vector (see below on how to choose which) until the coefficient of one of the vectors $\mbf v_l^*$ in the basis goes to zero, using the equation:
\bea
x=\mbox{max}_{a_k}: \forall_i b_i= (A^{-1} \mbf t)_i-a_k (A^{-1} \mbf v_k)_i\geq 0
\eea
\item The vector $\mbf v_k$ trades places with $\mbf v_l^*$.
\item We update the functional and the cost, which has decreased by $x\times RC$.
\item Repeat this procedure until we can no longer lower the cost.
\end{itemize}
{\tt JuliBootS} has currently two ways of deciding which vector to choose when there are multiple ones with negative reduced cost. These can be chosen as an option in {\tt iterate!}:
\bec
julia> iterate!(prob,n,method->"mrc")\\
julia> iterate!(prob,n,method->"mcv")
\eec
If no option is specified, the default method is ``mcv'' -- which stands for {\em maximum cost variation}. The MRC method selects the overall vector for which the reduced cost is the smallest -- for which \ref{negrc} is the most negative. But there is another ingredient to determine the total variation in cost, which is the maximum value $x$ of $a_k$ as determined by equation (\ref{xval}). For each vector with negative reduced cost there will be an associated $x$. The method MCV then selects a set of vectors with small reduced cost, and  chooses the one for which the total cost variation will be maximal. In practice we have found that, although MCV introduces an overhead per iteration of the simplex method relative to MRC, this is compensated by a greatly reduced number of overall interations. 
\section{Minimization}
\label{min}

In the previous section we saw how the simplex method proceeds by systematically modifying the vectors in the current solution in such a way as to reduce the cost. An important part in this procedure is to find a vector such that the functional is negative when acting on it, which in {\tt JuliBootS} is implemented by the {\tt findMRC} (and related) methods. Sitting behind this apparently simple function is in fact a large part of the {\tt JuliBootS} code which we would like to describe.

At every step of the simplex algorithm, we have a linear functional and a set of vectors in our search space. The search space is made up of a discrete set of vectors (an array of {\tt LPVector} types) and a discrete set of vector {\em functions} (an array of {\tt LPVectorFunc} types). The search for a vector with negative reduced cost is done for each of these. For the discrete set of vectors, the task is straightforward: we merely act with the functional on each of the vectors and pick the ones for which the result is negative. For vector functions, we may also dot the functional into each of them, but in this case we recover for each some (non-vector) function. This function of a single argument, $f(\Delta)$ represents the action of the functional on the vector obtained by evaluating the vector functions at $\Delta$. We want then to find, for each of these functions, values of $\Delta$ where they are negative. More than this, we are particularly interested in values of $\Delta$ which are local minima, since it will be for these that the reduced cost is smallest. Hence we are faced with a minimization problem.

A particular minimization algorithm is implemented in the {\tt bb} module. The link to this module is made in the {\tt LPLinks} file, and may be replaced in the future if a better method is developed. The algorithm has been described in detail in \cite{El-Showk2014a}. The idea is quite simple. Essentially we want to find a set of intervals such that in each of them the first derivative of the function $f(\Delta)$ is well approximated\footnote{``Well approximated'' is quantified by the parameter {\tt BB\_ISGOOD} in the {\tt consts.jl} file.} by a quadratic polynomial. Once we have found such intervals we then apply Newton's method in the ones which have the possibility of having a minimum. 

The determination of the intervals themselves is made using a standard iterative algorithm. We begin with the full range where the function is defined, and compare the exact value of the function at the center of the interval with approximations obtained by using second-order Taylor expansions at the ends of the interval. If the difference is small enough we are done. If not we divide the interval in two and repeat the procedure for each of them. We can continue in this fashion until we have good approximations for every interval.

\begin{figure}[htbp]
	\centering
		\includegraphics[width=10cm]{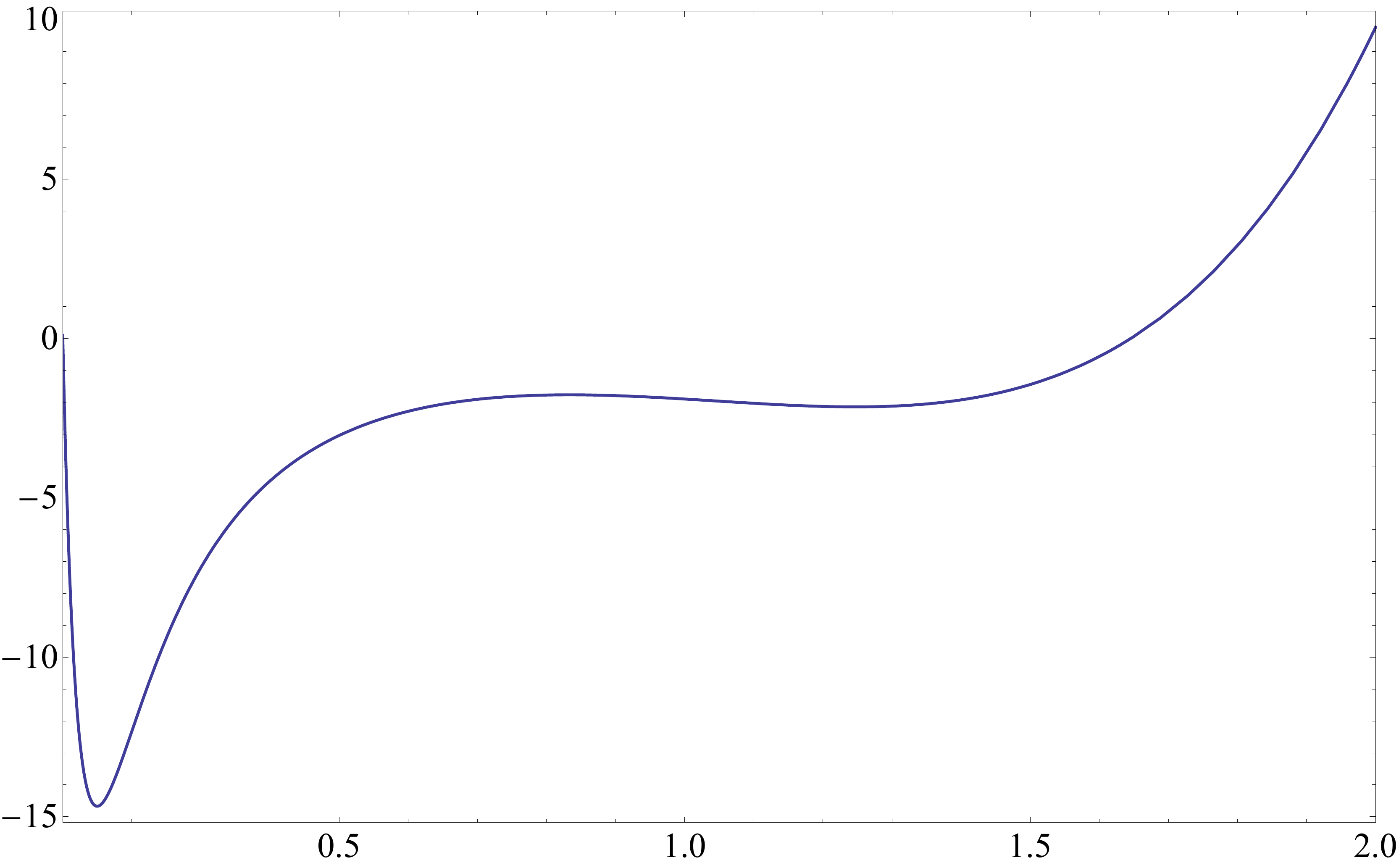}
	\label{fig:testfunction}
\end{figure}

Let us try out this in {\tt JuliBootS}. We begin by loading the package:
\bec
julia> using juliboots
\eec
Functions to be minimized are placed into {\tt MinFunction} types. These require knowledge not only of the function, but of its first, second and third derivatives. As an example, let us pick a rational function. Rational functions are implemented in the {\tt QFunc} module. Suppose we want to minimize the function
\bea
f(x)=x-3x^3+x^5-\frac{1}{x^2}+\frac{1/10}{x^3}
\eea
whose plot is shown in figure \ref{fig:testfunction}.
The rational function is made up of a polynomial and two poles:
\bec
julia> poly=1.0*Polynomial([0,1,0,-3,0,1]);\\
julia> pole1=Pole(2,0.,-1.);\\
julia> pole2=Pole(3,0.,0.1);\\
julia> f=poly+pole1+pole2
\eec
We now create a {\tt MinFunction} object from it:
\bec
julia> mf=MinFunction((0.1,2.),f)
\eec
The definition includes a range over which the function will be minimized. Finally we call the minimization routine
\bec
julia> FindMinimum(mf)\\
julia> (0.14979844483736832,-14.674944647035959)
\eec
This returns a doublet with the location and value of the global minimum. If we wish to find all local minima, we use instead
\bec
julia> FindLocalMinima(mf)
\eec
which now correctly returns a list containing the two local minima of this function. The accuracy in the determination of the minima can be controlled by adjusting parameters in the {\tt consts.jl} file.
\section{Bootstrapping: applications and examples}
\label{bs}
This section acts as a practical guide for solving typical problems in the conformal bootstrap. In the process we outline the main applications of {\tt JuliBootS} and its usage.

\subsection{Bounds with no global symmetry}
We begin by considering the simplest case described in the introduction: the correlator of four identical scalars $\sigma$ with dimension $\Delta_{\sigma}$. We wish to derive an upper bound on the dimension of the lowest scalar appearing in the OPE of $\phi$ with itself. To do this, we assume this bound lies within some initial range $(l,u)$. We then impose a gap in the scalar spectrum, from the unitarity bound up to the center of the range $c=(l+u)/2$, and look for solutions to the truncated crossing equations. If a solution is found we redefine $l=c$; if not then $u=c$; and repeat the process. In this way we can determine the upper bound with any desired accuracy. This procedure is called {\em bissection}.

Let us see how to do this in {\tt JuliBootS}. After launching {\tt Julia} on the {\tt JuliBootS} package directory, we load the code itself:\\
\begin{code}
julia> using juliboots
\end{code}
Next we set up the linear program that is to be solved by the simplex method. This involves loading a table of conformal blocks; transforming them into ``convolved blocks'', which incorporate the dependence on the dimension $\Delta_{\sigma}$; and setting up the initial Linear Program object. All this can be achieved with the function {\tt setupLP}, 
\begin{code}
julia> sigma=BigFloat(0.5182)\\
julia> prob=setupLP(sigma,"./Tables/D3\_n3m1\_L20.txt")
\end{code}
The function receives two arguments. The first is the value of the external dimension, here taken to be 0.5182 (which is the approximate dimension of the 3d Ising model spin field $\sigma$). The second is a conformal block table. A few tables are available in the GitHub repository, and more can be generated using the {\em Mathematica} notebook {\tt ComputeTables.nb} provided with {\tt JuliBootS}. These tables contain all the information required to generate the rational function representations of conformal blocks for a specific spacetime dimension, up to some maximum number of spins, and for a certain number of derivatives. Here we have loaded one of the tables available for spacetime dimension $d=3$, which give vectors with ten convolved components. The output of {\tt setupLP} is a {\tt LinearProgram}-type object.
\bec
julia> typeof(prob)\\
LinearProgram\{BigFloat\} (constructor with 1 method)\\
julia> prob\\

Linear Program:Basic Bound\\
D = 3.0	sigma=0.518	    (m,n) = (1,3)  	Lmax = 20	 Odd spins: false
\eec
The short description tells us that the derivatives used have $(m_{\mbox{\tiny max}},n_{\mbox{\tiny max}})=(1,3)$ -- this is explained in \ref{cb}, but essentially higher $m,n$ mean more derivatives, larger vectors and stronger results. Another important fact is that only blocks with (even) spins up to 20 are included. For the number of derivatives in this table this is more than sufficient to rule out truncation effects  discussed in section \ref{intro}.

The linear program contains a list of vector functions, one for each spin, from which the solver will attempt to build a solution to crossing symmetry. For instance,
\bec
julia> prob.lpFunctions[2]\\
LPVectorFunc - L=2, range - (3e+00 with 212 bits of precision,7e+01 with 212 bits of precision)
\eec
is the vector function associated with spin 2 blocks. It is defined over the range $\Delta \in (3,70)$. The lower limit is the unitarity bound, whereas the upper is the cutoff mentioned in section \ref{intro} -- a parameter which may be changed in the {\tt consts.jl} file, or by directly resetting the {\tt range} element in the corresponding {\tt lpFunction}. Again, for this low number of derivatives this number is more than enough. The full list of vector functions may be obtained with the command
\bec
julia> prob.lpFunctions
\eec
Given this Linear Program we are ready to derive bounds on lowest dimension operators. This is achieved with the {\tt bissect} command:
\bec
julia> upper=3\\
julia> lower=1\\
julia> accuracy=0.001\\
julia> type="L=0"\\
julia> result=bissect(prob,upper,lower,accuracy,type)
\eec
The command takes as arguments a linear program, upper, lower and accuracy arguments for the bissection procedure, and the type of object to bound. In this case bissection will be performed in the scalar (``L=0'') sector between 1 and 3 (since we expect the bound to be saturated by the Ising model operator $\varepsilon$ with dimension $\simeq 1.42$), and an upper bound will be determined with an absolute accuracy of 0.001.
During the calculation, information will be displayed on the current bissection point, number of iterations in the simplex method, time elapsed and current cost. The output of {\tt bissect} is a pair of new linear programs, which describe the output of the simplex method immediately below and above the upper bound. The current status of each Linear Program may be investigated with the {\tt status} function:
\bec
julia> status(result[1])
\eec
The result is a list of vectors currently used in the solution of the problem. If no AUX-type vectors exist, we have a genuine solution to crossing, and the total cost is zero. The reader can check that this is not the case for {\tt result[2]}. The upper bound is by definition the lowest allowed dimension in the Linear Program which could {\em not} be solved
\bec
julia> result[2].lpFunctions[1].range[1]
\eec
All in all, the full code required for running a basic bound is
\bec
julia> using juliboots\\
julia> prob=setupLP(0.518,"./Tables/D3\_n3m1\_L20.txt")\\
julia> bissect(prob,3,1,0.01,"L=0")
\eec
The reader should feel free to experiment with the parameters. For convenience one may want to save the results of a given run, namely the set of vectors that make up the solution to crossing. To do this a method is provided,
\bec
saveresults({\em file::String}, \em prob::LinearProgram)
\eec
This outputs the spectrum of a linear program {\tt prob} to {\tt file} in a {\em Mathematica} friendly format.

\subsection{Including vectors in the sum rule}

Suppose we want to force a specific block with a determined OPE coefficient to appear in the crossing symmetry sum rule. This could be useful {\em e.g.} instance if we are looking for CFTs with specific central charges, since the central charge fixes the OPE coefficient of the stress-tensor. A natural thing to consider in this case is to include the stress-tensor into the sum rule with a fixed coefficient and place an upper bound on the next spin-2 operator in the spectrum. Notice that if we do not place a gap in the spin-2 sector, nothing stops the solver from including a vector into the solution arbitrarily close to the stress-tensor (effectively increasing the OPE coefficient we start with). Depending on the gap we will get different results, so the most universal thing to do is to consider an upper bound on the dimension of the next spin-2 operator, $T'$.

We begin by setting up a basic Linear Program:
\bec
julia> using juliboots\\
julia> prob=setupLP(0.518,"./Tables/D3\_n3m1\_L20.txt")
\eec
The {\em target} of the Linear Program is the right-hand side of the crossing symmetry sum rule. Currently this is simply (minus) the identity operator, which is the scalar of dimension zero. We can check this explicitly:
\bec
julia> id=value(prob.lpFunctions[1],0);\\
julia> convert(Float64,norm(id+prob.target))\\
0.0
\eec
To include the stress tensor we first create an {\tt LPVector} object that will be associated to it, and modify the target using an in-built routine:
\bec

julia> T=makeVector(prob.lpFunctions[2],3)\\
LPVector - (3e+00 with 212 bits of precision, "L=2"), cost - 0e+00\\

julia> opecoeff=0.08;\\
julia> changeTarget!(prob,-id-opecoeff*T.vector);
\eec
And with this we simply perform bissection in the $L=2$ sector.
\bec
julia> result=bissect(prob,7,3,0.001,"L=2")
\eec
Depending on the value of the OPE coefficient the upper bound will vary -- as we encourage the reader to check.
\subsection{Filtering}
{\tt JuliBootS} includes routines for including or excluding vectors in the search space. We begin as usual by setting up a basic linear program,
\bec
julia> using juliboots\\
julia> prob=setupLP(0.6,"./Tables/D3\_n3m1\_L20.txt")
\eec
The routine {\tt iterate!} implements the simplex method on a Linear Program. Currently our problem has no gaps at all; all possible vectors from the unitarity bound up to the cut-off are in the search space. Therefore, it has to be possible to solve it. Let us check this. First we back up the original problem and then we use {\tt iterate!} (which modifies the LP that it acts on) to try to reduce the cost to zero.
\bec
julia> tmp=mcopy(prob);\\
julia> iterate!(tmp,500)\\
Started at: 25/09/2014 10:57:48		 Initial Cost: 71.01747020453475\\
Min cost achieved\\
Linear Program:Basic Bound\\
D = 3.0	sigma=0.518	(m,n) = (1,3)	Lmax = 20	Odd spins: true
\eec
The minimum possible cost has been achieved. We can check this is actually zero,
\bec
julia> cost(tmp)\\
julia> Cost: 0e+00
\eec
or by using {\tt status(tmp)}, which also tells us the specific solution that was found.

So far, so boring. But we can impose constraints on the operators allowed in the sum rule and see if it's still possible to find a solution. This is essentially what {\tt bissect} does in an iterative way. To change the spectrum we can use the method {\tt filter}:
\bec
julia> noSpinTwo=filter(prob,10,"L=2")
\eec
For instance, here we have removed from the search space all spin-2 operators with dimensions ranging from the unitary bound up to ten. If we now attemp to to solve the problem, we shall find that there is no solution.

We can apply more general kinds of filters with
\bec
filter({\em prob::LinearProgram,range::(Real,Real),criteria::Function})
\eec
This command excludes from the search space of {\tt prob} all vectors whose dimensions lie in {\tt range}, and whose label satisfies {\tt criteria(label)==True}. Typically this will create a new vector function in the search space, {\em e.g.} if we filter out the range (5,6) from an existing vector function defined  from (1,100) we will get two, one in (1,5) and another in (6,100). One typically uses this method to figure out in conjunction with {\tt iterate!} or {\tt bissect} to check if it is still possible to find a solution given this filtered out search space. In fact, the {\tt bissect} method is in its essence a sequence filter, check for solution, filter, check for solution, etc.

\subsection{OPE maximization}

In previous examples we have only seen how to check for existence of solutions to crossing symmetry. If we are well inside the allowed region (below any bounds on operator dimensions, say), the solution is not unique. One way of determining a unique solution is to assign a cost function and minimize that cost. This is useful for instance when we wish to find solutions which maximize an OPE coefficient. By assigning a {\em negative} cost to the corresponding vector, the simplex method will automatically converge on the solution which has maximal OPE coefficient for that vector.
A practical application is to determine lower bounds on the central charge of the theory. In this case, we maximize the OPE coeffcient of the stress tensor, which is inversely proportional to the central charge. 

Starting with a {\tt LinearProgram} type, we maximize an OPE coefficient in two steps. First, we must construct a feasible solution, {\em i.e.} a solution to crossing which does not involve auxiliary vectors. During this step, the costs of all vectors are set to zero (except the auxiliary ones of course). The second step is the maximization itself. Here we remove the auxiliary vectors from the search space so that they do not reappear, and set a negative cost to the vector whose OPE coefficient (squared) we which to maximize. In practice, it is simplest to explicitly add this vector independently to the search space with the desired cost. At this point the simplex method can start again, since it is possible to reduce the cost by bringing this vector into the basis.

We shall do these steps one by one for practice, although {\tt JuliBootS} has a single method for all of them. We begin by finding a feasible solution in our usual example in three-dimensions:
\bec
julia> using juliboots\\
julia> prob=setupLP(0.50001,"./Tables/D3\_n3m1\_L20.txt")\\
julia> iterate!(prob,500)
\eec
After a few iterations the problem is feasible. If we try iterating further we are told the minimum cost has been achieved, and we can check it is zero (using {\tt cost(prob)}. We now explicitly remove the auxiliary vectors by using {\tt filter}:
\bec
julia> prob2=filter(prob,Inf,"AUX")
\eec
Auxiliary vectors don't have a dimension per say, but for convenience we say that they do -- the $i$-th vector has dimension $i$ by convention. Filtering out up to infinity removes all vectors of type ``AUX'' from the {\tt LinearProgram}.

The next is to add a vector with negative cost. Let us add the stress tensor. First we generate it by evaluating the spin-2 vector function at unitarity:
\bec
julia> T=makeVector(prob2.lpFunctions[2],3)\\
LPVector - (3e+00 with 212 bits of precision,"L=2"), cost - 0e+00
\eec
We set the cost of $T$ to be negative,
\bec
julia> T.cost=-BigFloat(1)
\eec
and add $T$ to the search space:
\bec
julia> push!(prob2.lpVectors,T)
\eec
At this point the simplex method may be applied again until the minimum cost has been achieved
\bec
julia> iterate!(prob2,500)
\eec
We can check that the stress-tensor is indeed part of the solution now, using either {\tt prob2.solVecs}, {\tt solution(prob2)} or {\tt status(prob2)}. Its OPE coefficient (squared) has been maximized -- adding more constraints (going to higher nmax) can only decrease this maximum, so this is a valid upper bound. In this case we get a coefficient roughly equal to 0.081732. For technical reasons, this is not the true OPE coefficient squared that multiplies conformal blocks. To obtain the physical OPE coefficients\footnote{Recall that the physical normalization is chosen such that a four point function takes the form $\sum \lambda^2_{\Delta,L} G_{\Delta,L}(z,\bar z)$ with $G_{\Delta,L}\simeq (z\bar z)^{\Delta/2}$ for small $z=\bar z$.} we should correct by an extra factor:
\bea
\lambda_{\Delta,L}^2=(\lambda_{\Delta,L}^{JB})^2/(4\rho)^{\Delta}=(\lambda_{\Delta,L}^{JB})^2/(12-8\sqrt{2})^{\Delta}
\eea
This is because currently the code works only with the rational function part of the conformal blocks, omitting the $(4\rho)^{\Delta}$ prefactor in (\ref{derg}). In this case $\Delta=3$ and we get $\lambda_T^2\simeq 0.253$. To convert the OPE coefficient squared (which is the output of the code) into a central charge one can use the formula
\bea
C_T=\frac{d}{d-1}\, \frac{\Delta_{\sigma}^2}{\lambda_T^2}.
\eea
The normalization of $C_T$ here is such that for a free theory of $n_b$ bosons and $n_f$ Dirac fermions one gets
\bea
C_T=\frac{d}{d-1}\, n_b+d \,n_f
\eea
Plugging in $\Delta_\sigma$ and the OPE coefficient, setting $d=3$ we obtain $C_T=1.48309$. This sets a lower bound on all possible solutions to crossing symmetry. Since we've set $\Delta_\sigma$ at the free value, the result should be close to the free value too and indeed it is -- the central charge value is 3/2=1.5 for free theory in $d=3$.

Finally, we may avoid this sequence of steps by using the command 
\bec
opemax({\em lp::LinearProgram,dimension::Real,label}).
\eec
In our example we would've done
\bec
julia> using juliboots\\
julia> prob=setupLP(0.50001,"./Tables/D3\_n3m1\_L20.txt")\\
julia> res=opemax(prob,3,"L=2")
\eec
The output is the solved {\tt LinearProgram} with OPE (squared) coefficient maximized.

\subsection{Global symmetry}
\label{globalsym}
The {\tt JuliBootS} package supports applications to bootstrap correlators of fields charged under some global symmetry. We show how to do this in a concrete example. Consider the four-point function of a fundamental of $SO(N)$. The OPE of a fundamental with itself includes contributions from different tensor structures \cite{Rattazzi:2010yc}, namely singlets ($S^+$), traceless symmetric $T^+$ and antisymmetric $A^-$:
\bea
\phi_i \times \phi_j \simeq \sum_{S^+} \delta_{ij} \mathcal O^S+\sum_{T^+} \mathcal O^T_{(i,j)}+\sum_{A^-}\, \mathcal O^A_{[i,j]}
\eea
Accordingly there are three distinct bootstrap equations, which we may write as
\bea
\sum_{S^+}\lambda^2_{S}
\left(\begin{tabular}{c}
$0$\\
$F^{\df}_{\Delta,L}$\\
$H^{\df}_{\Delta,L}$
\end{tabular}
\right)
+
\sum_{T^+}\lambda^2_{T}
\left(\begin{tabular}{c}
$F^{\df}_{\Delta,L}$\\
$\left(1-\frac 2N\right)F^{\df}_{\Delta,L}$\\
$-\left(1+\frac 2N\right)H^{\df}_{\Delta,L}$
\end{tabular}
\right)
+
\sum_{A^-}\lambda^2_{A}
\left(\begin{tabular}{c}
$-F^{\df}_{\Delta,L}$\\
$F^{\df}_{\Delta,L}$\\
$-H^{\df}_{\Delta,L}$
\end{tabular}
\right)=0.
\eea
Here we have defined
\bea
F^{\df}_{\Delta,L}=G_{\Delta,L}(u,v)- u^{\Delta_\sigma} G_{\Delta,L}(v,u), \\
H^{\df}_{\Delta,L}=G_{\Delta,L}(u,v)+u^{\Delta_\sigma} G_{\Delta,L}(v,u). 
\eea
All three sums include all operators with dimension consistent with the unitarity bound, but the first two include only even spins, and the third odd spins.

To set up this problem in {\tt JuliBootS} we first need to define the three different kinds of vectors, their contents, and give them a name. Let us do this for $SO(3)$ say. Then:
\bec
julia> using juliboots\\
julia> N=3\\
julia> v1=[(1,"Z"),(1,"F""),(1,"H"),"even","S"]\\
julia> v2=[(1,"F"),(1-2/N,"F"),(-(1+2/N),"H"),"even","T"]\\
julia> v3=[(-1,"F"),(1,"F"),(-1,"H"),"odd","A"]\\
julia> vectortypes=(v1,v2,v3)
\eec
Here $v1,v2,v3$ stand for the three vectors in the sum rule. We specify their make up in terms of type -- $F$,$H$, or $Z$ (which stands for zero) -- coefficient, which kind of spins are allowed (odd, even or all) and finally a label for the vector -- in this case $S$,$T$ and $A$. Now we simply include the specification of the vector types into the {\tt setupLP} function which we have encountered before:
\bec
julia> sig=0.52;\\
julia> prob=setupLP(sig,"./Tables/D3\_n3m1\_L20\_allspins.txt",vectortypes)
\eec
and with this we are pretty much done. Now that we have our Linear Program we can bissect, do OPE maximization, etc, in much the same way as in previous examples. The only added complication is that vector functions (and vectors more generally) are labelled not only in terms of their spin, but also by the extra label defined in vector types. So, for instance we can derive a bound on the singlet sector by
\bec
julia> result=bissect(prob,3,1,0.01,"L=0 - S")
\eec
While we have shown the procedure for $SO(N)$, it is straightforward to generalize this example to more complicated cases. There is however an important caveat, which is that by making composite vector types we are effectively increasing the number of components of each vector -- for $SO(N)$ say, by a factor of 3. The larger number of components is similar to increasing the number of derivatives, and the larger number of vector types is analogous to increasing the total number of spins. The larger search space implies that calculations will therefore take longer.

\subsection{Scans and parallelization}

Usually we are interested in obtaining result not for a single value of $\Delta_\sigma$ but for a whole range. One way of doing is of course to do a loop, but this can be very time-consuming for higher numbers of derivatives. Here we shall mention two possible solutions. 

\subsubsection*{Using {\tt Julia}}

We begin by describing {\tt Julia}'s parallelization functions (for further details we advise the consultation of the {\tt Julia} manual at {\tt http://docs.julialang.org/en/release-0.3/}. In this first approach, the number of jobs does not have to match the number of threads available. This is convenient when running jobs on a single machine. 
The first thing to do is to launch multiple instances of {\tt Julia} and load \jb on each of them
\bec
julia> addprocs(2);\\
julia> @everywhere using juliboots;
\eec
The command {\tt @everywhere} is a macro, a special class of methods which are applied on {\tt Julia} expressions. In this case the macro performs the command on every available thread, in this case the master and the two slave threads added by {\tt addprocs}. We proceed by setting up a small set of linear programs
\bec
julia> sigs=[0.501:0.01:0.52];\\
julia> probs=setupLP(sigs,"./Tables/D3\_n3m1\_L20.txt");
\eec
To solve the problems in parallel we can use {\tt pmap}, which automatically parallelizes the application of a function to an array in parallel:
\bec
julia> tmp(x)=iterate!(x,500);\\
julia> res=pmap(tmp,probs)
\eec
The output is a list with the solved Linear Programs:
\bec
julia> map(cost,res)\\
3-element Array{Any,1}\\
0e+00\\
0e+00\\
0e+00
\eec
It is clear that this can be generalized for more complicated temporary functions which bissect, store results, etc.

\subsubsection*{Using {\tt JuliBootS}}
On a cluster we have typically many cores available and then it makes sense to allocate one core per job -- value of $\Delta_\sigma$. Included with {\tt JuliBoots} is a number of functions to help achieve this, contained in the {\tt Cluster} folder. 
Once things are properly set-up, the flow of a run is very simple: one first edits the {\tt specs.jl} file, which contains all the required specifications of a run. If required one can also adjust the {\tt JuliBootS} parameters in the {\tt consts.jl} file. In {\tt specs.jl} we determine the range of $\Delta_\sigma$ that will be sweeped, how accurate bissection will be performed, etc. To start the run we simply run on the command line:
\bec
julia quickstart.jl
\eec
This will setup a run directory as indicated in {\tt specs.jl}; launch a {\tt Julia} thread for each point; and save the results of the run using method {\tt saveresults} -- which outputs the spectrum at the bound. As this makes clear, currently the code is setup to do bissection but this is easy to change. The function that will be run by each core is determined by the {\tt runner.jl} file, which can be modified.

Before launching a run however, the most important thing is to indicate the cluster specific command that launches a job on the system. The cluster commands are specified in the file {\tt batchfuncs.jl}. In particular, there is currently a {\tt cernbatch} command adapted for the particular cluster at CERN. This method should be adapted to the user's particular case.

\acknowledgments

During this work I was supported by US DOE-grant DE-SC0010010 and by a Marie Curie Intra-European Fellowship of the
European Community's 7th Framework Programme under contract number PIEF-GA-2013-
623606.
I would like thank J. Golden, D. Poland, D. Simmons-Duffin and A. Vichi for discussions and collaboration, and special thanks go to S. El-Showk and V. Rychkov  for granting me access to their Python-based bootstrap code and for introducing me to {\tt Julia}.

\bibliography{Biblio}{}
\bibliographystyle{JHEP}
\end{document}